\input{aipcheck}

\documentclass[
  draft            
  ]
  {aipproc}

\usepackage{footmisc}
\usepackage{floatflt}

\layoutstyle{6x9}

\begin{document}
\title[Probing Extra Dimensions with ATLAS]
{Probing Extra Dimensions with ATLAS\footnote{~to 
appear in Proceedings of SUSY06, 
the 14th International Conference on Supersymmetry and the Unification of Fundamental Interactions, 
UC Irvine, California, 12-17 June 2006.}}


\author{Dominik Dannheim}{
  address={Columbia University, New York;
now at CERN, Experimental Physics Division,\\ CH-1211 Geneva 23, 
  Switzerland, Email: dominik.dannheim@cern.ch}}

\begin{abstract}
In the late nineties several authors suggested that the extra dimensions
predicted by string theory might lead to observable effects at high
energy colliders.  The ATLAS experiment which will start taking data at
the LHC in 2007 will be an excellent place to search for such effects.
The sensitivity of ATLAS to signatures of Extra Dimensions will
be presented.

\end{abstract}

\maketitle


\section{Introduction}

Over the past decade, new models based on large compactified
extra spatial dimensions
(ED) have been proposed~\cite{add,prl-83-1999-3370}, which could explain the large gap 
between the electroweak- (EW) and the Planck scale of 
$M_{EW}/M_{Pl}\approx 10^{-17}$. The $1/r$ Newtonian 
potential seen at large distances 
becomes a $1/r^{1+n}$ potential at small distances, implying a large 
fundamental Newton constant when the compactification radius is large.
Present direct limits on ED allow, for example, 2 ED up to 
0.13~mm~\cite{pr-d-70-042004-2004}.
For such distances, SM particles are constrained to reside on 
3 spatial dimensions by gauge theories of EW and strong interactions, 
valid to < fm distances, but gravitons could propagate in the bulk.
The resulting fundamental
scale of gravity could be as small as a few TeV and the
observed Planck mass would not be a relevant physical scale anymore.
A theoretical
foundation for introducing ED is given in the 
framework of string theories, which need ED
for self-coherence reasons. 

The LHC, colliding protons at a centre-of-mass energy of
$\sqrt{s}=14$~TeV, will exceed the accessible mass reach of the 
current collider experiments by an order of magnitude into the TeV 
range. The ATLAS detector~\cite{atlas-tdr} at the LHC, 
which will become operational 
in 2007, will therefore be an ideal place to look for ED signatures,
such as jets with missing
$E_T$, high-mass resonances or events with a large
$\Sigma E_T$. 
In the following, studies
are presented which evaluate the sensitivity
of ATLAS to detect signals from various ED scenarios. All results
are based on a parameterised simulation model of the ATLAS 
detector~\cite{atlfast}. Unless otherwise noted, an integrated
luminosity of $\int L$dt~$=100$~fb$^{-1}$ is assumed, corresponding to one year
of data taking at design luminosity.

\section{Large Extra Dimensions}

In the ADD model~\cite{add}, proposed by Arkani-Hamed, Dimopoulos and Dvali,
the SM fields are confined to a 4D brane. 
A new gravity scale $M_D\approx$TeV is
introduced in 4+n dimensions. In the 4D brane, the graviton
appears as a tower of closely spaced (<1~meV)
massive Kaluza-Klein (KK) excitation states
with universal coupling to the SM fields. The coupling is small
($\approx1/M_{Pl}$), but nevertheless a sizeable cross section is obtained from the
large number of interfering KK states. Virtual exchange of KK states
leads to deviations in cross sections at high mass or energy and deviations in 
angular distributions.
Direct production of gravitons results in jet + missing $E_T$ 
signatures. 

ATLAS has evaluated the sensitivity to detect virtual graviton exchange in 
deviations in the dilepton and diphoton invariant mass 
spectrum~\cite{atl-phys-2001-012}. The enhancement in the SM cross sections
is obtained as function of an effective scale $M_S\approx M_D$. 
For this 
scenario, ATLAS will be able to probe ED up to $M_S\approx
7-8$~TeV.

Direct graviton production with the graviton escaping undetected in the 
ED has been studied by searching for a possible excess of
events with jets and missing energy~\cite{jpg-27-2001-1839}. 
In this scenario,
ED can be probed up to $M_D \approx 6-9$~TeV.

A variation of the ADD scenario was also considered~\cite{epj-c-2004-10-1140},
in which additional
small ED orthogonal to the brane are present with a 
compactification radius $M_c\approx $TeV$^{-1}$. Fermions are confined to 
the 3-brane, but gauge fields propagate in the
TeV$^{-1}$-scale ED. Leptonic decays of gauge excitations
of the photon and the $Z$ boson would provide striking signatures
at the LHC. The sensitivity of ATLAS in this scenario has been estimated
to reach to values of $M_c\approx 5.8$~TeV.
A study of the angular distribution of the produced lepton pairs 
in the peak region showed
that $\int L$dt~$=300$~fb$^{-1}$
would be needed to distinguish
such gauge excitations from mass peaks produced by
SM-like $Z'$ or graviton resonances 
(for $M_c<5$~TeV). 

\section{Warped Extra Dimensions}

Graviton resonances are predicted in several models with small ED.
In a Randall-Sundrum model~\cite{prl-83-1999-3370},
the metric is not flat, but varies exponentially in an extra dimension 
between the TeV brane, on which SM particles reside, and the Planck scale.
This "warped" dimension is responsible
for the shrinkage of the gravitational interaction strength in
our (3+1)D subspace. The boundary conditions in the ED
give rise to periodic wave functions; the KK modes in (3+1)D for the 
excited graviton $G^*$. The mass of the first resonance is given
as 
$m_1\approx3.83 \Lambda_{\pi} k/\bar{M}_{Pl}$,~with~$0.01 < k/\bar{M}_{Pl}
< 0.1$.
The cross section for $G^*$ production is proportional to $k/\bar{M}_{Pl}^2$.
The couplings of the $G^*$ are universal, giving rise to many decay 
channels, such as $ee$, jet-jet, $\gamma \gamma$, $ZZ$, $WW$.
The spin-2 nature of the $G^*$ can be used in a spin analysis~\cite{jhep-11-2005-046}
to distinguish it from SM background and from other exotic scenarios,
such as $Z'$ production.

\begin{figure} \label{graviton_limit}
\includegraphics[height=.25\textheight]{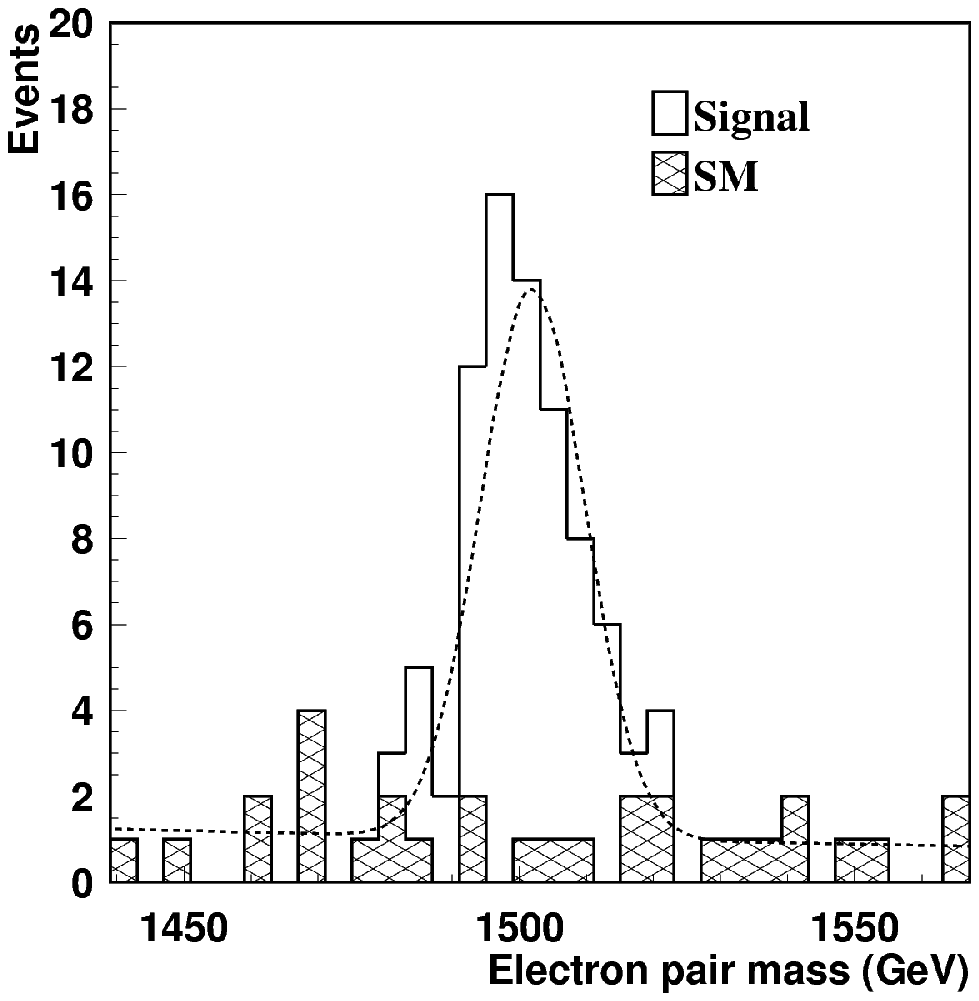}
\hspace{0.5cm}
\includegraphics[height=.25\textheight]{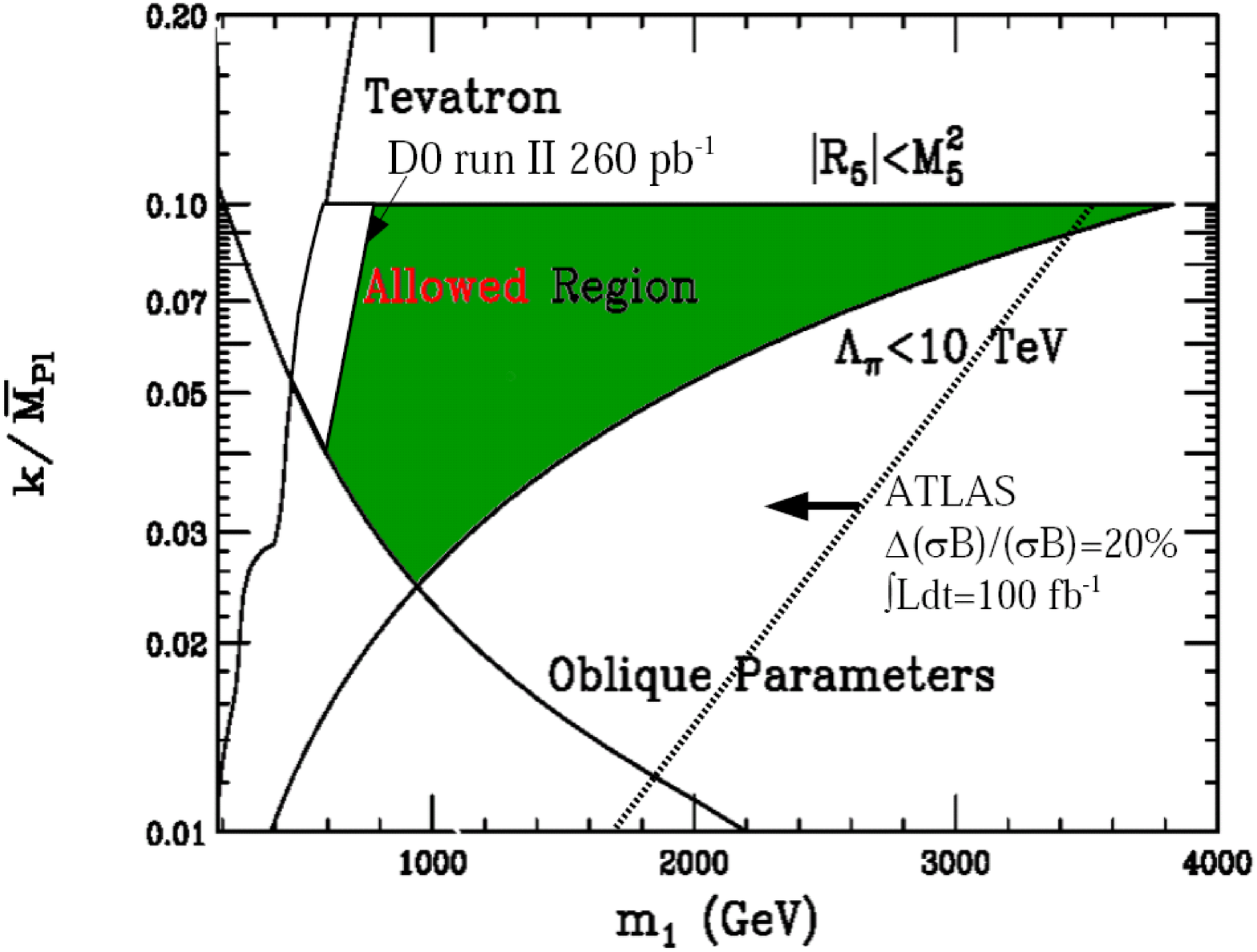}
\begin{picture}(0,0) 
\put(-395, 145){a)} \put(-205, 145){b)}
\put(-358, 135){\small{$\int L$dt~$=100$~fb$^{-1}$}}
\end{picture}
\caption{a) Invariant mass distribution for RS graviton-production at $M_G$=1.5~TeV
and for the SM background. b) Current constraints on the RS parameter space. The expected
sensitivity of ATLAS is indicated with a dashed line.}
\end{figure}

The sensitivity for $G^*$ detection was studied in several of the possible
decay channels~\cite{allanach-graviton}.
The decay to two electrons is considered the most promising discovery
channel. It has a low branching ratio 
of $\approx 2\%$, but a very clear signature of 2 back-to-back high 
$p_T$ electrons with an invariant mass equal to the mass of the
$G^*$. The background from SM processes is small, mostly originating
from the high-mass tail of the DY spectrum. 
Figure \ref{graviton_limit}a) shows the 
two-electron invariant mass spectrum in ATLAS consisting of
SM background and a hypothetical graviton resonance at 1.5~TeV 
in a scenario with $k/\bar{M}_{Pl}=0.01$. The signal
is clearly visible above the low SM background.
The ultimate $5\sigma$
discovery reach for $k/\bar{M}_{Pl}=0.01$ was 
estimated to be 2080~GeV, well above the current experimental limit
of 250~GeV from the the D0 experiment at Run II of the Tevatron~\cite{prl-95-2005-091801}. 
The currently allowed region of the
RS parameter space is shown in Fig. \ref{graviton_limit}b), together with the ATLAS reach
for a determination of the $G^*$-production cross section with an
accuracy of 20\%. 
The interesting region of the parameter space is
covered by ATLAS.

\section{Universal Extra Dimensions}

Universal Extra Dimension (UED) scenarios allow all SM particles
to propagate in ED of TeV$^{-1}$ size. Due to 
tree-level KK-number conservation the KK 0-mode states can not be 
virtual and can only be produced in pairs. 
There are only weak
constraints from previous electro-weak measurements. 
The current
Tevatron results constrain the mass of the compactification
scale to $M_c>400$~GeV~\cite{prd-64-035002-2001}. 

For ATLAS, a study for a 
specific class of UED models has been performed~\cite{atl-phys-pub-2005-03}. It assumes that the topology of the
UED model forms a thick brane embedded in a bulk of two large
ED, where only gravity propagates. Only models
with one ED of TeV$^{-1}$ size and only gravity-mediated
decays of the KK states have been considered. The signature in this
case are dijet events with large missing transverse energy. 
SM background
from events with jets and a $W$ or a $Z$ boson was considered. 
At $5\sigma$,
ATLAS will be sensitive to $M_c\approx 2.7$~TeV.

A recent phenomenological study~\cite{fermilab-pub-06-009-t} 
considered higher modes of KK excitations, which could be 
singly produced at colliders. They are heavier than the (1,0) modes
by a factor $\sqrt{2}$ and their cascade decays lead to a series
of narrow-spaced $t\bar{t}$ resonances, resulting in lepton + jets events.
Within this scenario, previous $t\bar{t}$ resonance studies have
been reinterpreted. A 1~TeV $t\bar{t}$ resonance from UED could be observed
with  $\int L$dt~$=30$~fb$^{-1}$. The reach could be further increased by
taking into account extra leptons and jets produced in the cascade 
decays. Furthermore, if the difference of $\sqrt{2}$ between consecutive
even states would be observed, one could discriminate from other
scenarios and extract the number of ED.

\section{Micro Black Holes}

At the LHC, micro black holes (BH) could be produced if the fundamental
Planck scale would be lowered to $\approx$TeV~\cite{pr-d-65-056010-2002,prl-87-2001-161602}.
A BH is characterised by its
mass, $M_{BH}=\sqrt{s}$, its Schwarzschild radius $R_S=2 G M_{BH}/c^2$
and (for $n$ ED) 
by its temperature $T_H\approx (1+n)/(M_{BH}^{1/(1+n)})$.
The production cross section at the LHC for BHs would be about
$\sigma\approx \pi R_S^2 \approx 100$~pb. BHs
decay with a short lifetime of $\approx 10^{-27} - 10^{-25}$~s via 
Hawking radiation to all degrees of freedom with equal probability.
Such decays will thus result in high-multiplicity, high $\Sigma p_T$
spherical events with significant missing $E_T$. The BH mass can be
reconstructed from the sum of all 4 momenta and the temperature can
be extracted from a comparison of the energy spectrum of the final-state
particles to the expectation from a black-body spectrum.

BH studies for ATLAS have been performed with $n=2-6$ ED
and a BH mass from 5~TeV up to 14~TeV~\cite{epj-c-41-2005-19,jhep-05-2005-053}. Background from
QCD events at large missing $E_T$ and from SUSY has been considered.
It was shown that ATLAS will be able to discover higher-dimensional
BHs and to explore their parameters.
The BH parameters were extracted
in a test model with $n=4$ and $M_{BH}=7$~TeV and assuming that the 
parton-level BH-production cross section is known to 20\%. The accuracy
obtained in this case was about 10\% for the temperature, 15\% for the
mass and $\pm 0.75$ for the number of ED.

\section{Conclusions}

ATLAS will have an unprecedented reach for discovery of ED
in the TeV range. Moreover, discrimination between models and extraction
of model parameters will be possible in several scenarios.


\bibliographystyle{aipprocl} 


\begin{thebibliography}{9}

\bibitem{add}
N. Arkani-Hammed, S. Dimopoulos and G. Dvali, \emph{Phys. Lett.} \textbf{B429},
263 (1998); \emph{Phys. Rev.} \textbf{D59}, 086004 (1999).

\bibitem{prl-83-1999-3370}
L. Randall and R. Sundrum, \emph{Phys. Rev. Lett.} \textbf{83}, 3370 (1999);
\textbf{83}, 4690 (1999).

\bibitem{pr-d-70-042004-2004}
C.D. Hoyle et al., \emph{Phys. Rev.} \textbf{D70}, 042004 (2004).

\bibitem{atlas-tdr} 
ATLAS collaboration, ATLAS Technical Design Report, CERN/LHCC 99-14/15 (1999).

\bibitem{atlfast}
E. Richter-Was, D. Froidevaux and L. Poggioli, ATLAS Internal Note
ATL-PHYS-98,131 (1998).

\bibitem{atl-phys-2001-012}
V. Kabachenko, A. Miagkov and A. Zenin, ATLAS Internal Note ATL-PHYS-2001-12
(2001).

\bibitem{jpg-27-2001-1839}
L. Vacavant and I. Hinchliffe,
\emph{J. Phys.}, \textbf{G27} no. 8, 1839 (2001).

\bibitem{epj-c-2004-10-1140}
G. Azuelos and G. Polesello,
\emph{Eur. Phys. Journ.} \textbf{C}, 10.1140 (2004).

\bibitem{jhep-11-2005-046}
R. Cousins et al., \emph{J. High Energy Phys.} \textbf{11}, 046 (2005).

\bibitem{allanach-graviton}
B.C. Allanach et al., \emph{J. High Energy Phys.} \textbf{9}, 019 (2000);
\textbf{12}, 039 (2002).

\bibitem{prl-95-2005-091801}
V.M. Abazov et al., \emph{Phys. Rev. Lett.} \textbf{95}, 091801 (2005).

\bibitem{prd-64-035002-2001}
T. Appelquist, H.-C. Cheng and B.A. Dobrescu, \emph{Phys. Rev.} \textbf{D64},
035002 (2001).

\bibitem{atl-phys-pub-2005-03}
P.-H. Beauchemin and G. Azuelos, ATLAS Internal Note ATL-PHYS-PUB-2005-03 (2005).

\bibitem{fermilab-pub-06-009-t}
G. Burdman, B.A. Dobrescu and E. Ponton, FERMILAB-Pub-06-009-T (2006).

\bibitem{pr-d-65-056010-2002}
S. Giddings and S. Thomas, \emph{Phys. Rev.} \textbf{D65}, 056010 (2002).

\bibitem{prl-87-2001-161602}
S. Dimopoulos, G. Landsberg,
\emph{Phys. Rev. Lett.} \textbf{87}, 161602 (2001)

\bibitem{epj-c-41-2005-19}
J. Tanaka et al.,
\emph{Eur. Phys. J.} \textbf{C41}, 19 (2005)

\bibitem{jhep-05-2005-053}
C.M. Harris et al.,
\emph{J. High Energy Phys.} \textbf{05}, 053 (2005)

\end{thebibliography}

\end{document}